\documentclass[12pt,a4paper]{article}%
\usepackage{amssymb}
\usepackage{amsfonts}
\usepackage{amsmath}
\usepackage{graphicx}%
\providecommand{\U}[1]{\protect\rule{.1in}{.1in}}

\begin{document}

\title{\textbf{A finite set of equilibria for the indeterminacy of linear rational
expectations models}}
\author{Jean-Bernard Chatelain\thanks{Paris School of Economics, Universit\'{e} Paris
I Pantheon Sorbonne, CES, Centre d'Economie de la Sorbonne, 106-112 Boulevard
de l'H\^{o}pital 75647 Paris Cedex 13. Email:
jean-bernard.chatelain@univ-paris1.fr} and Kirsten Ralf\thanks{ESCE
International Business School, 10 rue Sextius Michel, 75015 Paris, Email:
Kirsten.Ralf@esce.fr.}}
\maketitle

\begin{abstract}
This paper demonstrates the existence of a finite set of equilibria in the
case of the indeterminacy of linear rational expectations models. The number
of equilibria corresponds to the number of ways to select $n$ eigenvectors
among a larger set of eigenvectors related to stable eigenvalues. A finite set
of equilibria is a substitute to continuous (uncountable) sets of sunspots
equilibria, when the number of independent eigenvectors for each stable
eigenvalue is equal to one.

\textbf{JEL\ classification numbers}: C60, C61, C62, E13, E60.


\end{abstract}

\begin{quotation}
\textit{"Das kann als Riccatische gleichung des matrizenkalk\"{u}ls angesehen
werden." }Radon (1928) p.190.
\end{quotation}

\section{Introduction}

This paper demonstrates that there is a finite set of rational expectations
equilibria in the case of indeterminacy for linear rational expectations
models, which is a \emph{substitute} to uncountable (continuously infinite)
sets of sunspots equilibria (Gourieroux et al. (1982)). This occurs when the
number of independent eigenvectors for each stable eigenvalue is equal to one,
in particular, when all stable eigenvalues are distinct. This paper extends
Blake and Kirsanova (2012) results for time-consistent optimal policy rules to
the general case of Blanchard Kahn (1980) solutions.

Blanchard and Kahn (1980) states that there are multiple equilibria with
rational expectations (or indeterminacy) when the number $n$ of pre-determined
variables is lower than the number $s$ of eigenvalues below one in absolute
values. In this case, the initial values of the number $m$ of non
pre-determined "forward" variables may be driven by continuous random
variables of zero mean, independently and identically distributed over time
(Gourieroux \textit{et al.} (1982)).

Besides this continuous infinity of sunspots equilibria, it is feasible to
extend the computation of saddlepath unique rational expectations equilibrium
(Blanchard and Kahn (1980), Boucekkine and Le Van (1996)) to the case of
multiple equilibria. These rational expectations equilibria are solutions of a
matrix Riccati equation (Radon (1928), Le Van (1986), Abou-Kandil \textit{et
al.} (2003)). This paper demonstrates that there is a finite a number of
equilibria, at most equal to $\frac{s!}{n!(s-n)!}$.\ This is the number of
ways to choose $n$ distinct eigenvectors among a larger set of $s$
eigenvectors related to eigenvalues with absolute values below one, when there
is only one independent eigenvector for each of these eigenvalues.

\section{A finite set of equilibria with indeterminacy}

Blanchard and Kahn (1980) consider a\ linear rational expectations model:%

\begin{equation}
\left(
\begin{array}
[c]{c}%
\mathbf{k}_{t+1}\\
_{t}\mathbf{q}_{t+1}%
\end{array}
\right)  =\underset{\mathbf{A}}{\underbrace{\left(
\begin{array}
[c]{cc}%
\mathbf{A}_{nn} & \mathbf{A}_{nm}\\
\mathbf{A}_{mn} & \mathbf{A}_{mm}%
\end{array}
\right)  }}\left(
\begin{array}
[c]{c}%
\mathbf{k}_{t}\\
\mathbf{q}_{t}%
\end{array}
\right)  +\mathbf{\gamma z}_{t}%
\end{equation}
where $\mathbf{k}_{t}$ is an $\left(  n\times1\right)  $ vector of variables
predetermined at $t$ with initial conditions $\mathbf{k}_{0}$ given (shocks
can straightforwardly be included into this vector); $\mathbf{q}$ is an
$\left(  m\times1\right)  $ vector of variables non-predetermined at $t$;
$\mathbf{z}$ is an $\left(  k\times1\right)  $ vector of exogenous variables;
$\mathbf{A}$ is $\left(  n+m\right)  \times\left(  n+m\right)  $ matrix,
$\mathbf{\gamma}$ is a $\left(  n+m\right)  \times k$ matrix, $_{t}%
\mathbf{q}_{t}$ is the agents expectations of $\mathbf{q}_{t+1}$ defined as follows:%

\begin{equation}
_{t}\mathbf{q}_{t+1}=E_{t}\left(  \mathbf{q}_{t+1}\shortmid\Omega_{t}\right)
.
\end{equation}
$\Omega_{t}$ is the information set at date $t$ (it includes past and current
values of all endogenous variables and may include future values of exogenous
variables). A \textbf{predetermined} variable is a function only of variables
known at date $t$ so that $\mathbf{k}_{t+1}=$ $_{t}\mathbf{k}_{t+1}$ whatever
the realization of the variables in $\Omega_{t+1}$. A
\textbf{non-predetermined} variable can be a function of any variable in
$\Omega_{t+1}$, so that we can conclude that $\mathbf{q}_{t+1}=$
$_{t}\mathbf{q}_{t+1}$ only if the realization of all variables in
$\Omega_{t+1}$ are equal to their expectations conditional on $\Omega_{t}$.

Boundary conditions for the policy-maker's first order conditions are the
given initial conditions for predetermined variables $\mathbf{k}_{0}$ and
Blanchard and Kahn (1980) hypothesis ruling out the exponential growth of the
expectations of $\mathbf{w=}\left(  \mathbf{k,q,z}\right)  $:%

\begin{equation}
\forall t\in%
\mathbb{N}
\text{,}\exists\overline{\mathbf{w}}_{t}\in%
\mathbb{R}
^{k}\text{,}\exists\theta_{t}\in%
\mathbb{R}
\text{, such that }\left\vert E_{t}\left(  \mathbf{w}_{t+1}\shortmid\Omega
_{t}\right)  \right\vert \leq\left(  1+i\right)  ^{\theta_{t}}\overline
{\mathbf{w}}_{t}\text{, }\forall i\in%
\mathbb{R}
^{+}.
\end{equation}
\qquad

\textbf{Definition:} Besides other sunspots equilibria (Gourieroux et al.
[1982]), let us define a set of rational expectations solutions, which are
such that non predetermined variables are a linear function of pre-determined
variables, where the matrix $\mathbf{N}_{mn}$ is to be found, and with bounded
solutions for pre-determined variables, so that the eigenvalues $\lambda_{i}$
of the matrix $\mathbf{A}_{nn}-\mathbf{A}_{nm}\mathbf{N}_{mn}$ are below one
("stable eigenvalues"):%

\begin{align}
\mathbf{q}_{t+1}  &  =-\mathbf{N}_{mn}\mathbf{k}_{t+1}\\
\mathbf{k}_{t+1}  &  =\left(  \mathbf{A}_{nn}-\mathbf{A}_{nm}\mathbf{N}%
_{mn}\right)  \mathbf{k}_{t}\\
\lambda\left(  \mathbf{A}_{nn}-\mathbf{A}_{nm}\mathbf{N}_{mn}\right)   &
=\left\{  \lambda_{i}\text{ with }\left\vert \lambda_{i}\right\vert
<1,i\in\left\{  1,...,n\right\}  \right\}
\end{align}

\textbf{Proposition: }$\mathbf{A}$\textit{\ has }$s$\textit{\ stable
eigenvalues} \textit{and }$n+m-s$\textit{\ unstable eigenvalues. }

\textit{Case 1. When }$0\leq s<n$\textit{, the number of stable eigenvalues is
strictly below the number of pre-determined variables, there is no rational
expectations equilibrium (Blanchard and Kahn (1980)).}

\textit{Case 2. When }$s=n$\textit{, the number of stable eigenvalues is
strictly equal to the number of predetermined variables, there is a unique
rational expectations equilibrium (Blanchard and Kahn (1980)).}

\textit{Case 3. When }$n<s\leq n+m$\textit{, the number of rational
expectations equilibria defined above is given by the number of ways of
selecting }$n$\textit{\ independent (right column) eigenvectors }$\left(
\begin{array}
[c]{c}%
\mathbf{P}_{nn}\\
\mathbf{P}_{mn}%
\end{array}
\right)  $ \textit{among a larger set of independent eigenvectors related to
stable eigenvalues. If }$\mathbf{P}_{nn}$ \textit{is invertible, they
corresponds to the number of rational expectations equilibria determined by
each matrix }$\mathbf{N}_{mn}=-\mathbf{P}_{mn}\mathbf{P}_{nn}^{-1}$\textit{:}

\textit{Case 3.1. \textbf{Finite number of equilibria}. If the number of
independent eigenvectors (geometric multiplicity) of each stable eigenvalues
of }$\mathbf{A}$\textit{ is exactly one, the number of equilibria is given by
}$\frac{s_{1}!}{n!s_{1}!}$\textit{ where the number of stable eigenvalues not
counting their multiplicity is denoted }$s_{1}\leq s$\textit{. In particular,
if all the stable eigenvalues of }$\mathbf{A}$\textit{ are distinct, then the
number of equilibria is }$\frac{s!}{n!s!}$\textit{.}

\textit{Case 3.2. \textbf{Uncountable number of equilibria}. If there is at
least one stable eigenvalue of }$\mathbf{A}$\textit{ with its number of
independent eigenvectors (geometric multiplicity) which is at least equal to
two, then, there always exists an uncountable number of equilibria. This
condition for an uncountable number of equilibria is distinct from e.g.
Gourieroux et al. (1982). }

For example, for $n=1$, $m=1$, and with a unique stable eigenvalue
$\lambda_{1}$ with two independent column vectors\textit{ }$\mathbf{P}%
=(\mathbf{P}_{1},\mathbf{P}_{2})$\textit{, }there is an uncountable number of
single eigenvectors\textit{ }$\mathbf{P}_{\alpha}=\mathbf{P}_{1}%
+\alpha\mathbf{P}_{2}$\textit{ }with $\alpha\in%
\mathbb{C}
$ leading to solutions\textit{ }$\mathbf{N}_{mn,\alpha}=-\mathbf{P}%
_{mn,\alpha}\mathbf{P}_{nn,\alpha}^{-1}$\textit{. }For $n=2$, $m=1$, including
another eigenvalue $\lambda_{3}$ with a multiplicity equal to one and an
eigenvector denoted $\mathbf{P}_{3}$, there is a single case of $n=2$ columns
eigenvector $(\mathbf{P}_{1},\mathbf{P}_{2})$ and an uncountable number of
$n=2$ eigenvector matrix $\mathbf{P}_{3\alpha}=(\mathbf{P}_{3},\mathbf{P}%
_{\alpha})$ with $\alpha\in%
\mathbb{C}
$ allowing to compute solutions\textit{ }$\mathbf{N}_{mn}=-\mathbf{P}%
_{mn}\mathbf{P}_{nn}^{-1}$ (see a numerical example for $n=2$, $m=2$ in
Abou-Kandil \textit{et al.} (2003) p.25).

\textbf{Proof: }

Let us consider a matrix $\mathbf{N}_{mn}$ such that:%

\begin{align}
\left(
\begin{array}
[c]{c}%
\mathbf{k}_{\mathbf{N,}t}\\
\mathbf{q}_{\mathbf{N,}t}%
\end{array}
\right)   &  =\left(
\begin{array}
[c]{cc}%
\mathbf{I}_{n} & \mathbf{0}_{nm}\\
-\mathbf{N}_{mn} & \mathbf{I}_{m}%
\end{array}
\right)  \left(
\begin{array}
[c]{c}%
\mathbf{k}_{t}\\
\mathbf{q}_{t}%
\end{array}
\right) \nonumber\\
\text{ with }\mathbf{T}  &  \mathbf{=}\left(
\begin{array}
[c]{cc}%
\mathbf{I}_{n} & \mathbf{0}_{nm}\\
-\mathbf{N}_{mn} & \mathbf{I}_{m}%
\end{array}
\right)  \text{ and }\mathbf{T}^{-1}\mathbf{=}\left(
\begin{array}
[c]{cc}%
\mathbf{I}_{n} & \mathbf{0}_{nm}\\
\mathbf{N}_{mn} & \mathbf{I}_{m}%
\end{array}
\right)
\end{align}

So that:%
\begin{align}
\left(
\begin{array}
[c]{c}%
\mathbf{k}_{\mathbf{N,}t+1}\\
\mathbf{q}_{\mathbf{N,}t+1}%
\end{array}
\right)   &  =\left(
\begin{array}
[c]{cc}%
\mathbf{I}_{n} & \mathbf{0}_{nm}\\
\mathbf{N}_{mn} & \mathbf{I}_{m}%
\end{array}
\right)  \left(
\begin{array}
[c]{cc}%
\mathbf{A}_{nn} & \mathbf{A}_{nm}\\
\mathbf{A}_{mn} & \mathbf{A}_{mm}%
\end{array}
\right)  \left(
\begin{array}
[c]{cc}%
\mathbf{I}_{n} & \mathbf{0}_{nm}\\
-\mathbf{N}_{mn} & \mathbf{I}_{m}%
\end{array}
\right)  \left(
\begin{array}
[c]{c}%
\mathbf{k}_{\mathbf{N,}t}\\
\mathbf{q}_{\mathbf{N,}t}%
\end{array}
\right) \nonumber\\
\left(
\begin{array}
[c]{c}%
\mathbf{k}_{\mathbf{N,}t+1}\\
\mathbf{q}_{\mathbf{N,}t+1}%
\end{array}
\right)   &  =\left(
\begin{array}
[c]{cc}%
\mathbf{A}_{nn}-\mathbf{A}_{nm}\mathbf{N}_{mn} & \mathbf{A}_{nm}\\
g(\mathbf{N}_{mn}) & \mathbf{A}_{mm}+\mathbf{N}_{mn}\mathbf{A}_{nm}%
\end{array}
\right)  \left(
\begin{array}
[c]{c}%
\mathbf{k}_{\mathbf{N,}t}\\
\mathbf{q}_{\mathbf{N,}t}%
\end{array}
\right)  \text{ with}\\
g(\mathbf{N}_{mn})  &  =\mathbf{A}_{mn}+\mathbf{A}_{mm}\mathbf{N}%
_{mn}-\mathbf{N}_{mn}\mathbf{A}_{mm}-\mathbf{N}_{mn}\mathbf{A}_{nm}%
\mathbf{N}_{mn}=\mathbf{0}_{mn}%
\end{align}

$g(\mathbf{N}_{mn})=\mathbf{0}_{mn}=\partial\mathbf{N}_{mn}/\partial t$ is a
matrix equation including a constant, two linear terms and a quadratic term
$\mathbf{N}_{mn}\mathbf{A}_{nm}\mathbf{N}_{mn}$, which Radon (1928) denoted as
matrix Riccati extension of scalar Riccati differential equations. If
$\mathbf{N}_{mn}$ is a solution with constant coefficients of $g(\mathbf{N}%
_{mn})=\mathbf{0}_{mn}$, then the characteristic polynomial of matrix
$\mathbf{A}$ is the product of two characteristic polynomials, as $\det\left(
\mathbf{T}\right)  =1=\det\left(  \mathbf{T}^{-1}\right)  $:
\begin{equation}
\det\left(  \mathbf{A-}\lambda\mathbf{I}_{n+m}\right)  =\det\left(
\mathbf{A}_{nn}-\mathbf{A}_{nm}\mathbf{N}_{mn}\mathbf{-}\lambda\mathbf{I}%
_{n}\right)  \cdot\det\left(  \mathbf{A}_{mm}+\mathbf{N}_{mn}\mathbf{A}%
_{nm}\mathbf{-}\lambda\mathbf{I}_{m}\right)  =0
\end{equation}

Each solution $\mathbf{N}_{mn}$ of $g(\mathbf{N}_{mn})=\mathbf{0}_{mn}$
corresponds to a particular partition of the eigenvalues of the matrix
$\mathbf{A}$ since its eigenvalues are exactly the eigenvalues of
$\mathbf{A}_{nn}-\mathbf{A}_{nm}\mathbf{N}_{mn}$ (with $n$ eigenvalues
counting multiplicity) and $\mathbf{A}_{mm}+\mathbf{N}_{mn}\mathbf{A}_{nm}$
(with $m$ eigenvalues counting multiplicity). A Jordan canonical
transformation $\mathbf{J}$ of the $\mathbf{A}$ matrix with $\mathbf{P}$ a
matrix of right eigenvectors is:%

\begin{equation}
\left(
\begin{array}
[c]{cc}%
\mathbf{A}_{nn} & \mathbf{A}_{nm}\\
\mathbf{A}_{mn} & \mathbf{A}_{mm}%
\end{array}
\right)  \left(
\begin{array}
[c]{cc}%
\mathbf{P}_{nn} & \mathbf{P}_{nm}\\
\mathbf{P}_{mn} & \mathbf{P}_{mm}%
\end{array}
\right)  =\left(
\begin{array}
[c]{cc}%
\mathbf{P}_{nn} & \mathbf{P}_{nm}\\
\mathbf{P}_{mn} & \mathbf{P}_{mm}%
\end{array}
\right)  \left(
\begin{array}
[c]{cc}%
\mathbf{J}_{n} & \mathbf{0}_{nm}\\
\mathbf{0}_{mn} & \mathbf{J}_{m}%
\end{array}
\right)
\end{equation}

where $\mathbf{J}_{nn}$\textbf{\ }is a $n\times n$ Jordan matrix with the
eigenvalues of $\mathbf{A}_{nn}-\mathbf{A}_{nm}\mathbf{N}_{mn}$ and
$\mathbf{J}_{mm}$ is a $m\times m$ Jordan matrix with the eigenvalues of
$\mathbf{A}_{mm}+\mathbf{N}_{mn}\mathbf{A}_{nm}$. One has:%

\begin{align}
&  \left(
\begin{array}
[c]{cc}%
\mathbf{A}_{nn}-\mathbf{A}_{nm}\mathbf{N}_{mn} & \mathbf{A}_{nm}\\
\mathbf{0}_{mn} & \mathbf{A}_{mm}+\mathbf{N}_{mn}\mathbf{A}_{nm}%
\end{array}
\right)  \left(
\begin{array}
[c]{cc}%
\mathbf{I}_{n} & \mathbf{0}_{nm}\\
-\mathbf{N}_{mn} & \mathbf{I}_{m}%
\end{array}
\right)  \left(
\begin{array}
[c]{cc}%
\mathbf{P}_{nn} & \mathbf{P}_{nm}\\
\mathbf{P}_{mn} & \mathbf{P}_{mm}%
\end{array}
\right) \nonumber\\
&  =\left(
\begin{array}
[c]{cc}%
\mathbf{I}_{n} & \mathbf{0}_{nm}\\
-\mathbf{N}_{mn} & \mathbf{I}_{m}%
\end{array}
\right)  \left(
\begin{array}
[c]{cc}%
\mathbf{P}_{nn} & \mathbf{P}_{nm}\\
\mathbf{P}_{mn} & \mathbf{P}_{mm}%
\end{array}
\right)  \left(
\begin{array}
[c]{cc}%
\mathbf{J}_{n} & \mathbf{0}_{nm}\\
\mathbf{0}_{mn} & \mathbf{J}_{m}%
\end{array}
\right)
\end{align}

which implies:%

\begin{equation}
\left(
\begin{array}
[c]{cc}%
\left(  \mathbf{A}_{nn}-\mathbf{A}_{nm}\mathbf{N}_{mn}\right)  \mathbf{P}%
_{nn}+\mathbf{A}_{nm}\left(  \mathbf{P}_{mn}-\mathbf{N}_{mn}\mathbf{P}%
_{nn}\right)  & \ast\\
\left(  \mathbf{A}_{mm}+\mathbf{N}_{mn}\mathbf{A}_{nm}\right)  \left(
\mathbf{P}_{mn}-\mathbf{N}_{mn}\mathbf{P}_{nn}\right)  & \ast
\end{array}
\right)  =\left(
\begin{array}
[c]{cc}%
\mathbf{P}_{nn}\mathbf{J}_{nn} & \ast\\
\left(  \mathbf{P}_{mn}-\mathbf{N}_{mn}\mathbf{P}_{nn}\right)  \mathbf{J}_{nn}
& \ast
\end{array}
\right)
\end{equation}

Because the eigenvalues of $\mathbf{A}_{mm}+\mathbf{N}_{mn}\mathbf{A}_{nm}$
are not the eigenvalues of $\mathbf{J}_{nn}$, then $\left(  \mathbf{P}%
_{mn}-\mathbf{N}_{mn}\mathbf{P}_{nn}\right)  $ cannot stack eigenvectors (each
of them distinct from the zero vector by definition) of $\mathbf{A}%
_{mm}+\mathbf{N}_{mn}\mathbf{A}_{nm}$. Then, the second equality for block
matrices $(i=2,j=1)$ is valid:%

\begin{equation}
\left(  \mathbf{A}_{mm}+\mathbf{N}_{mn}\mathbf{A}_{nm}\right)  \left(
\mathbf{P}_{mn}-\mathbf{N}_{mn}\mathbf{P}_{nn}\right)  =\left(  \mathbf{P}%
_{mn}-\mathbf{N}_{mn}\mathbf{P}_{nn}\right)  \mathbf{J}_{nn}%
\end{equation}

only and only if $\mathbf{P}_{mn}-\mathbf{N}_{mn}\mathbf{P}_{nn}=\mathbf{0}$.
Then, if $\mathbf{P}_{nn}$ is invertible, one finds the solutions
$\mathbf{N}_{mn}=$ $-\mathbf{P}_{mn}\mathbf{P}_{nn}^{-1}.$

According to the rational expectations equilibria definition, one needs to
find at least $n$ stable eigenvalues, and compute $\mathbf{N}_{mn}$ using a
set of $n$ column eigenvectors $\left(
\begin{array}
[c]{c}%
\mathbf{P}_{nn}\\
\mathbf{P}_{mn}%
\end{array}
\right)  $ related to these stable eigenvalues. The number of rational
expectations equilibria is then given by the number of ways of selecting
$n$\ independent (right column) eigenvectors $\left(
\begin{array}
[c]{c}%
\mathbf{P}_{nn}\\
\mathbf{P}_{mn}%
\end{array}
\right)  $ related to the stable eigenvalues $s\geq n$.

Finally, the first equality for block matrices $(i=1,j=1)$ becomes:%

\begin{equation}
\left(  \mathbf{A}_{nn}-\mathbf{A}_{nm}\mathbf{N}_{mn}\right)  \mathbf{P}%
_{nn}=\mathbf{P}_{nn}\mathbf{J}_{nn}%
\end{equation}

Hence, the matrix $\mathbf{P}_{nn}$ is an eigenvectors matrix of the matrix
$\mathbf{A}_{nn}-\mathbf{A}_{nm}\mathbf{N}_{mn}$. \textbf{Q.E.D.}

A similar demonstration with transpose matrices holds for left row
eigenvectors $\left(
\begin{array}
[c]{c}%
\mathbf{Q}_{mn}\\
\mathbf{Q}_{mm}%
\end{array}
\right)  $ with $\mathbf{Q}=\mathbf{P}^{-1}$ chosen among a set of $s>n$\ row
eigenvectors related to stable eigenvalues. If $\mathbf{Q}_{mm}$ is
invertible, one finds the solutions $\mathbf{N}_{mn}=-\mathbf{P}%
_{mn}\mathbf{P}_{nn}^{-1}=\mathbf{Q}_{mm}^{-1}\mathbf{Q}_{mn}$.

\section{Conclusion}

A finite set of rational expectations equilibria (when the number of
independent eigenvectors for each stable eigenvalue is equal to one) exists at
each period. For a chosen equilibrium with a given set of eigenvectors at a
given period to be find again on the following periods, one needs to assume
that the economic agents select the same set of eigenvectors at each period.
In this case, economic agents shape their rational expectations following the
same procedure at each period in a time-consistent manner.

\end{document}